\documentclass[prd,aps,nofootinbib,preprint,eqsecnum,tightenlines]{revtex4}
\usepackage{xcolor}
 \usepackage{rotating}
\usepackage{array}
\usepackage{amsmath,amssymb}
\usepackage{graphicx}
\usepackage{mathrsfs}
\usepackage{bm}
\usepackage{dcolumn}
\usepackage{float}
\usepackage{placeins}
\usepackage{makecell}
\usepackage{tabularx}
\usepackage{enumerate}
\usepackage{hhline}
\usepackage{array}
\usepackage{tabularx}
\usepackage{graphicx}
\usepackage{pdflscape}
\usepackage{makecell}

\begin{document}
	
\title{Self-consistent Hubble expansion in exponential teleparallel gravity: confrontation with recent observations}
\author{K.~Sri~Kavya$^{1}$, T.~Vinutha$^{2}$, B.~Revathi$^{2}$, Kazuharu~Bamba$^{3}$}

\affiliation{$^{1}$Department of Science and Humanities, MVGR College of Engineering, Vizianagaram, India.}

\affiliation{$^{2}$Department of Applied Mathematics, Andhra University, Visakhapatnam, India,}

\affiliation{$^{3}$Faculty of Symbiotic Systems Science, Fukushima University, Fukushima, Japan.}

	\begin{abstract}
We reconstruct the Hubble expansion history within an anisotropic exponential teleparallel gravity framework using recent cosmological observations, including Observational Hubble Data (OHD), the latest Dark Energy Spectroscopic Instrument Data Release 2 (DESI DR2) baryon acoustic oscillation (BAO) measurements, gravitational-wave (GW) standard-siren data, and the Pantheon Plus and SH0ES Type Ia supernova compilations. The proposed anisotropic exponential $f(T)$ model is investigated in both the absence and presence of shear, and is confronted with these independent observational probes through statistical analyses based on the reduced $\chi^{2}$, Akaike Information Criterion (AIC), and Bayesian Information Criterion (BIC), demonstrating its consistency with current cosmological observations. Furthermore, the cosmological viability of the model is investigated through a comprehensive study of the dynamical evolution, stability conditions, and linear matter perturbations, providing a unified assessment of its theoretical consistency and observational compatibility.

	\end{abstract}
	
	\maketitle
	\newpage
	\section{Introduction}
	The expansion of the universe has remained an intriguing research topic in recent decades, but several astronomers have elucidated that the universe comprises  two fundamental constituents: namely, dark matter and dark energy. These enigmatic components significantly influence the evolution of the universe. Two eminent research teams \cite{Riess:1998,Perlmutter:1999} have concluded that dark energy plays a pivotal role in the study of the universe's accelerated expansion \cite{Luca:2015}. Moreover, various observational datasets have provided indirect yet compelling evidence of this cosmic acceleration.Several theoretical frameworks have been proposed to explain this cosmic acceleration, one of which is Teleparallel Gravity (TG) \cite{Cai:2016}, which is equivalent to General Relativity (GR). 
	
	Teleparallel Gravity (TG) is a dynamically equivalent reformulation of General Relativity (GR), where gravity is described by spacetime torsion rather than curvature. In contrast to GR, which is based on curved spacetime geometry, TG is formulated as a gauge theory of translations in a flat spacetime with torsion. Although the theories differ conceptually, they yield the same gravitational field equations; therefore, TG is commonly known as the Teleparallel Equivalent of General Relativity (TEGR) \cite{Mirza:2017}.
	Among the modified theories of gravity, $f(R)$ gravity has been extensively studied as a generalisation of GR, in which the Einstein--Hilbert action is extended by replacing the Ricci scalar $R$ with a general function $f(R)$. Such modifications provide a geometrical approach to explain the early-time inflationary epoch and the late-time accelerated expansion of the Universe without introducing additional exotic matter components \cite{DeFelice:2010aj,Sotiriou:2008rp,Clifton:2011jh,Nojiri:2010wj,Nojiri:2017ncd,Capozziello2011,Bamba:2012cp,Odintsov:2026a,Odintsov:2026b,Odintsov:2026c,Odintsov:2025}. Motivated by the success of $f(R)$ gravity, one of the most important extensions of TEGR is $f(T)$ gravity, in which the torsion scalar is replaced by a general function $f(T)$. Such models offer an alternative geometrical approach to describing the observed cosmic acceleration. In addition, teleparallel dark energy models with scalar--torsion couplings and non-local extensions of $f(T)$ gravity have also been investigated as viable alternatives in cosmology \cite{Geng:2011,Otalora:2013,Bahamonde:2017,Cai:2015}.
	
	
	The applications of $f(T)$ gravity have been extensively explored in both cosmological and gravitational studies. In particular, these models have been widely used to explain the late--time accelerated expansion of the universe without introducing a cosmological constant and can effectively mimic quintessence-like behaviour \cite{Cai:2016,Dent:2011}. Different functional forms, including power-law and exponential modifications of the torsion scalar, have also been tested using observational data from type Ia supernovae, baryon acoustic oscillations, cosmic microwave background radiation, and cosmic chronometers \cite{Bamba:2012,Bamba:2013,Santos:2022,Koussour:2024}. Beyond cosmology, $f(T)$ gravity has been applied to astrophysical and solar-system phenomena such as gravitational lensing, fast radio bursts, planetary motion, and light deflection \cite{Jiang:2024,Xie:2013}. Torsion-based corrections have furthermore been explored in the context of inflationary and early-universe cosmology \cite{Dimakis:2024,Ghosh:2025}. The theoretical foundations and covariant formulation of teleparallel gravity are discussed in Refs.~\cite{Golovnev:2017,Bahamonde:2021gfp}, while viable cosmological applications of $f(T)$ gravity have been extensively studied in Ref.~\cite{Bamba:2010wb}. Overall, these developments demonstrate the importance of $f(T)$ gravity as a promising framework for cosmology and gravitational physics.

  Anisotropy refers to the directional dependence of physical properties in space or matter.
	In cosmology, an anisotropic universe expands at different rates along distinct spatial directions,
	in contrast to the isotropic case, where the expansion is uniform.
	Observationally, anisotropies manifest through temperature fluctuations in the Cosmic Microwave Background (CMB),
	hemispherical power asymmetry, and preferred directional patterns in the large-scale structure of the universe.
	Investigating such features provides critical tests of the cosmological principle of large-scale isotropy
	and places constraints on models of the early universe, inflationary scenarios,
	and possible deviations from the standard $\Lambda$CDM cosmology. 
	
	 Despite extensive studies, most investigations of $f(T)$ gravity have been confined to isotropic and homogeneous Friedmann--Lema\^itre--Robertson--Walker (FLRW) geometries.
	However, several recent observations, including CMB hemispherical asymmetry and large-scale dipole anisotropies \cite{Secrest:2021}, suggest that mild deviations from isotropy may exist on cosmological scales.
	These findings have motivated renewed interest in anisotropic cosmological models as potential extensions of the standard $\Lambda$CDM paradigm.
	In particular, the Bianchi type-I spacetime provides a simple and physically significant framework to investigate direction-dependent expansion rates and the isotropization process at late times.
	Several authors have carried out detailed studies of anisotropic cosmological models in different contexts (see Refs.~\cite{Akarsu:2019, Vinutha:2021, Vinutha:2020, Vinutha:2022, Koivisto:2008, Saha:2006, Rodrigues:2012}). 	
Previous studies of exponential $f(T)$ gravity have predominantly focused on isotropic FLRW cosmology. In contrast, the present work investigates the cosmological viability of exponential $f(T)$ gravity within an anisotropic Bianchi type-I spacetime using recent OHD, DESI DR2 BAO, Pantheon Plus, Pantheon+SH0ES, and gravitational-wave standard-siren observations. The primary objective is to determine whether the observed late-time cosmic acceleration can be consistently reproduced in the physically viable regime $|\nu T| \ll 1$. Furthermore, by performing a direct comparison between cosmological evolutions with and without shear contributions, we assess the role of cosmic anisotropy, late-time isotropization, and torsional modifications in shaping the expansion history of the universe.

Combining recent cosmological observations with dynamical, stability, and perturbation analyses, we show that the proposed model provides a physically viable description of late-time cosmic evolution while supporting late-time isotropization. In addition, the cosmological viability of the model is examined through dynamical system analysis, stability conditions, and linear matter perturbations. In Sec.~III, we explore the Bianchi type-I cosmological model and the corresponding field equations. In Sec.~IV, we describe the observational datasets and statistical methodology. In Sec.~V, we summarize our analyses and discuss the results. Finally, conclusions are given in Sec.~VI.
 
	\section{General mathematical formulation of $f(T)$ gravity}
	 Motivated by the analytical simplicity and geometrical structure of $f(T)$ gravity, we consider a generalized teleparallel framework in which the torsion scalar $T$ appearing in the Teleparallel Equivalent of General Relativity (TEGR) action is generalized to an arbitrary differentiable function $f(T)$. This generalization preserves the second-order structure of the resulting field equations, unlike the fourth-order equations encountered in $f(R)$ gravity, thereby rendering the theory both theoretically compelling and computationally efficient. In this investigation, we adopt the conventional action for $f(T)$ gravity, expressed as
	\begin{equation}\label{g1}
		S = \frac{1}{16 \pi G} \int f(T) \, |e| \, d^{4}x + \mathcal{L}_\mathrm{m} \, |e| \, d^{4}x,
	\end{equation}
	where $G$ denotes the gravitational constant, and $|e|$ denotes the absolute value of the determinant of the tetrad$\big(e^{A}_{i}\big)$. 
	The term $\mathcal{L}_\mathrm{m}$ corresponds to the matter Lagrangian density. For analytical convenience, the general function $f(T)$ is often written in the form
	\begin{equation}\label{g2}
		f(T) = T + F(T),
	\end{equation}
	where $T$ is the torsion scalar and $F(T)$ encodes deviations from the Teleparallel Equivalent of General Relativity (TEGR). In $f(T)$ gravity, the torsion scalar is defined by
	\begin{equation}\label{g3}
		T = T^{\gamma}{}_{ij} \, S^{ij}{}_{\gamma},
	\end{equation}
	where $T^{\gamma}{}_{ij}$ and $S^{ij}{}_{\gamma}$ denote the torsion tensor and the superpotential respectively. The torsion tensor is given by
	\begin{equation}\label{g4}
		T^{\gamma}{}_{ij} = e^{\gamma}{}_{A} \left( \partial_{i} e^{A}{}_{j} - \partial_{j} e^{A}{}_{i} \right),
	\end{equation}
	while the superpotential takes the form
	\begin{equation}\label{g5}
		S^{ij}{}_{\gamma} = \frac{1}{2} \left( K^{ij}{}_{\gamma} + \delta^{i}_{\gamma} \, T^{\alpha j}{}_{\alpha} - \delta^{j}_{\gamma} \, T^{\alpha i}{}_{\alpha} \right).
	\end{equation}
Here, $K^{ij}{}_{\gamma}$ is the contorsion tensor, defined by
	\begin{equation}
		\label{g6}K^{ij}{}_{\gamma} = \frac{1}{2} \left( T^{ji}{}_{\gamma} + T_{\gamma}{}^{ij} - T^{ij}{}_{\gamma} \right).
	\end{equation}
Varying the action \eqref{g1} with respect to the  $e^{A}_{i}$
	yields the field equations in $f(T)$ gravity as 
	\begin{equation}\label{g7}
		\begin{aligned}
			e^{-1} \, \partial_{i} \left( e \, e^{\gamma}{}_{A} \, S_{\gamma}{}^{ij} \right) \left(1+f_{T}\right) 
			- e^{\lambda}{}_{A} \, T^{\gamma}{}_{i\lambda} \, S_{\gamma}{}^{ji}\left(1+f_{T}\right)
			+ e^{\gamma}{}_{A} \, S_{\gamma}{}^{ij} \, \partial_{i} T \, f_{TT}
			\\&\hskip -5.5cm - \frac{1}{4} e^{j}{}_{A} \, \left[T+f(T)\right]
			= 4 \pi G \, e^{\gamma}{}_{A} \, T_{\gamma}{}^{j}.
		\end{aligned}
	\end{equation}
	Within this framework, $f_{T}$ and $f_{TT}$ correspond to the first and second order functional derivatives of $f(T)$ with respect to the torsion scalar respectively. $\mathbf{T}_{\gamma}{}^{j}$ denotes the stress-energy–momentum tensor derived from the matter field Lagrangian.
	
	In the present work, we consider an exponential form of $f(T)$ gravity given by
	\begin{equation}
		\label{g8}
		f(T)=T+\mu e^{\nu T},
	\end{equation}
	where $\mu$ and $\nu$ are arbitrary constants characterizing deviations from the Teleparallel Equivalent of General Relativity (TEGR). The parameter $\mu$ determines the strength of the exponential correction, while $\nu$ governs the evolution rate of the torsional modification.
	
	Exponential forms of $f(T)$ gravity have attracted considerable attention in modified teleparallel cosmology due to their ability to provide viable late--time cosmic accelerationwhile maintaining second-order gravitational dynamics \cite{Linder:2010py,Bamba:2011pz}. In particular, exponential corrections can generate small deviations from the standard $\Lambda$CDM cosmology and can successfully describe the observed accelerated expansion of the universe without introducing an explicit cosmological constant. Furthermore, such models can remain compatible with observational constraints and stability conditions for suitable choices of the model parameters \cite{Cai:2016}.
	
	The exponential form considered in the present work naturally reduces to the TEGR limit when the correction term becomes sufficiently small. Therefore, the model provides a theoretically consistent and observationally viable framework for investigating anisotropic cosmological evolution in teleparallel gravity. The dynamical analysis, stability conditions, and linear perturbation behaviour of the present model are discussed in Secs.~V.D -Sec.~ V.F.
\section{Model and field equations}
In standard cosmology, the large–scale structure of the universe is described by the Friedmann–Lemaître–Robertson–Walker (FLRW) metric, which assumes spatial homogeneity and isotropy. The corresponding FLRW spacetime metric takes the form
\begin{equation}
	\label{g9}
	ds^2 = dt^2 - a^2(t)\sum_{i=1}^{3} (dx^i)^2,
\end{equation}
 where $a(t)$ denotes the cosmological scale factor that determines how spatial distances evolve with cosmic time. The corresponding expansion rate is described by the Hubble parameter, defined as $H = \frac{\dot{a}}{a}$,
where the overdot represents differentiation with respect to cosmic time $t$. The discovery of the universe’s accelerated expansion has significantly redefined modern cosmology, introducing dark energy as a dominant yet elusive component of the cosmic framework. Although the standard $\Lambda$CDM model presumes large-scale isotropy and homogeneity, there is no compelling theoretical or observational evidence to ensure that such conditions persisted throughout cosmic history. Indeed, observations from Type Ia supernovae, the Cosmic Microwave Background (CMB), and large-scale structure indicate that the universe may have undergone anisotropic phases, especially during its early evolutionary stages.

This motivates the exploration of anisotropic cosmological models, which provide a natural generalisation of the FLRW framework by incorporating directional dependence in the dynamics of cosmic expansion. Such models gain particular relevance when coupled with dynamical dark energy components, such as scalar fields, as they facilitate a more comprehensive understanding of cosmic acceleration, structure formation, and potential departures from the cosmological principle. Consequently, the investigation of anisotropic cosmologies incorporating dark energy is both theoretically well-founded and observationally significant.

Among such models, the Bianchi type-I spacetime represents the simplest homogeneous but anisotropic generalisation of the FLRW metric. It is characterized by a diagonal metric with three independent directional scale factors, allowing different expansion rates along mutually orthogonal spatial directions.
Motivated by these considerations, in this article, we consider a time-dependent Bianchi type-I spacetime of the form
\begin{equation}
	\label{g10}
	ds^2 = dt^2 - a_{1}^2(t)\,dx^2 - a_{2}^2(t)\,dy^2 - a_{3}^2(t)\,dz^2,
\end{equation}
where $a_1(t)$, $a_2(t)$, and $a_3(t)$ are directional scale factors depending only on cosmic time $t$. Several researchers have investigated this spacetime in various cosmological contexts (see, Ref.~\cite{Sharif:2009, Koussour:2022}).
 
 The vierbein corresponding to the Bianchi type-I metric in Eq.~\eqref{g10} is chosen as
 \begin{equation}
 \label{g11}	e^A_{\ \mu} = \mathrm{diag}\big(1,\; a_1(t),\; a_2(t),\; a_3(t)\big).
 \end{equation}
 
 Using this tetrad, the torsion scalar for the Bianchi type-I spacetime is obtained as
\begin{equation}
 	\label{g12}
 	T = -2 \left( H_x H_y + H_y H_z + H_z H_x \right),
 \end{equation}
 where $H_x = \frac{\dot{a}_1}{a_1}$, $H_y = \frac{\dot{a}_2}{a_2}$, and $H_z = \frac{\dot{a}_3}{a_3}$ are the directional Hubble parameters and the mean Hubble parameter is given by
 \begin{equation}
 \label{g13}	H = \frac{1}{3}(H_x + H_y + H_z).
 \end{equation}
 In anisotropic cosmological models such as Bianchi type-I, the matter sector may, in general, exhibit directional dependence in pressure along the independent spatial directions. To account for this feature in a compact manner, the matter distribution can be described by an effective anisotropic fluid energy--momentum tensor of the form
\begin{equation}
	\label{g14}
	T^{\mu}{}_{\nu} = \mathrm{diag}\left(\rho_m,\,-p_x,\,-p_y,\,-p_z\right),
\end{equation}
where $\rho_m$ denotes the matter energy density, and $p_x$, $p_y$, and $p_z$ represent the pressures along the $x$, $y$, and $z$ directions, respectively.

 With the aid of Eqs.~\eqref{g5}, \eqref{g6}, \eqref{g7}, \eqref{g11},  \eqref{g12}, and \eqref{g14} the $f(T)$ field equations for the metric \eqref{g10} are obtained as
\begin{equation}
	\label{g15}
	\hskip -4.5 cm 2\left(H_xH_y + H_yH_z + H_zH_x\right) f_T + \frac{1}{2} f(T) = 8\pi \rho_m,
\end{equation}
\begin{equation}
	\label{g16_new}
	\left(\dot{H}_y + \dot{H}_z + H_y^2 + H_z^2 + H_yH_z\right) f_T 
	+ \frac{1}{2} f(T)
	+ (H_y + H_z)\dot{T} f_{TT}
	= -8\pi p_x,
\end{equation}
\begin{equation}
	\label{g17_new}
	\left(\dot{H}_x + \dot{H}_z + H_x^2 + H_z^2 + H_xH_z\right) f_T 
	+ \frac{1}{2} f(T)
	+ (H_x + H_z)\dot{T} f_{TT}
	= -8\pi p_y,
\end{equation}
\begin{equation}
	\label{g18_new} \hskip 1.1cm
	\left(\dot{H}_x + \dot{H}_y + H_x^2 + H_y^2 + H_xH_y\right) f_T 
	+ \frac{1}{2} f(T)
	+ (H_x + H_y)\dot{T} f_{TT}
	= -8\pi p_z.
\end{equation}

 The shear scalar quantifies the anisotropy in the expansion rates of the universe and is given by
 \begin{equation}
 \label{g19}	\sigma^2 = \frac{1}{6}\left[(H_x - H_y)^2 + (H_y - H_z)^2 + (H_z - H_x)^2\right].
 \end{equation}
 
  Using Eqs.~\eqref{g8}, \eqref{g12}, \eqref{g15}, and \eqref{g19}, the Hubble parameter for the metric \eqref{g10} is given by
\begin{equation} \label{20}
	H(z) = H_0\sqrt{
		\Omega_{\sigma 0} (1+z)^6
	+\frac{\Omega_{m0}(1+z)^3}{1+\mu\nu}
		 - \frac{\mu}{6(1 + \mu \nu)H_0^2} 
	},
\end{equation}
where $\Omega_{m0}=\frac{8\pi \rho_{m0}}{3 H_0^2}$ and $\Omega_{\sigma0}=\frac{{\sigma_0}^2}{3 H_0^2}$, denote the present-day matter and shear density parameters, respectively.
The above expression for $H(z)$ is obtained in the regime $|\nu T| \ll 1$. 
	
The derived Hubble parameter explicitly incorporates the anisotropic shear contribution through the term $\Omega_{\sigma0}(1+z)^6$, while naturally reducing to the corresponding non-shear cosmology in the limit $\Omega_{\sigma0}\rightarrow0$. This formulation enables a direct comparison between cosmological evolutions with and without shear within the exponential $f(T)$ framework and provides the basis for investigating the observational consequences of anisotropy using recent cosmological datasets.
 \section{Observational Data and Methodological Framework}
In this section, we present a comprehensive description of the observational data sets employed in the present analysis. The discussion highlights the sources, nature, and significance of the data, ensuring clarity regarding their role in constraining the model parameters. Particular emphasis is placed on the relevance of these data sets for testing the theoretical framework and establishing consistency with current cosmological observations.
\subsection{Observational Hubble Data (OHD)}
Direct measurements of the Hubble parameter $H(z)$ from the
Observational Hubble Data (OHD), which consist of observational
measurements of the Hubble expansion rate at different redshifts, provide
a model-independent probe of the cosmic expansion history. In this work,
we utilize a compilation of 33 $H(z)$ measurements spanning the redshift
range $0.07 < z < 2.36$, of which 31 are obtained using the cosmic
chronometer (CC) method. This method estimates $H(z)$ through differential
age measurements of passively evolving galaxies at nearby redshifts, as
first proposed in Ref.~\cite{Jimenez:2003} and subsequently refined in
Refs.~\cite{Stern:2010,Moresco:2012}. Further improvements and extensions
to higher redshifts were presented in
Refs.~\cite{Zhang:2014,Ratsimbazafy:2017,Moresco:2020}. In addition, two
high-redshift data points at $z=2.34$ and $z=2.36$ are derived from BAO
measurements in the Ly$\alpha$ forest \cite{Delubac:2015}. The combined
dataset provides robust constraints on the cosmic expansion rate over a
wide redshift range.
The present cosmological model is constrained using the independent
parameter vector
\begin{equation}
	\Theta=
	(H_0,\Omega_{m0},\Omega_{\sigma0},\nu).
\end{equation}
Here, $\Omega_{\sigma0}$ represents the present-day shear contribution,
while $\mu$ is determined from the normalization condition $E(0)=1$
and is therefore treated as a derived parameter rather than an
independent free parameter.
The corresponding chi-square statistic for the OHD dataset is given by
\begin{equation}
	\chi^2_{\mathrm{OHD}}
	=
	\sum_{k=1}^{33}
	\frac{
		\left[
		H^{\mathrm{obs}}(z_k)
		-
		H^{\mathrm{th}}(z_k,\boldsymbol{\Theta})
		\right]^2
	}
	{
		\sigma_H^2(z_k)
	}.
\end{equation}
Here, $H^{\mathrm{obs}}(z_k)$ and $\sigma_H(z_k)$ represent the observed value of the Hubble parameter and its associated uncertainty at redshift $z_k$, respectively, while $H^{\mathrm{th}}(z_k,\boldsymbol{\Theta})$ denotes the theoretical prediction of the model corresponding to the parameter vector $\boldsymbol{\Theta}$.

\subsection{BAO}
Baryon Acoustic Oscillations (BAO) provide one of the most powerful observational probes of the late-time expansion history of the Universe. The characteristic acoustic scale imprinted in the large-scale distribution of galaxies serves as a standard ruler for measuring cosmological distances and constraining the expansion history. In this work, we employ the latest BAO measurements from the Dark Energy Spectroscopic Instrument (DESI) Data Release 2 (DR2) \cite{Karim:2025}, which provide precise observational constraints through the distance indicators $D_M/r_d$, $D_H/r_d$, and $D_V/r_d$ over a broad redshift range. Here, $D_M$, $D_H$, and $D_V$ denote the transverse comoving distance, Hubble distance, and volume-averaged distance, respectively. The sound horizon at the drag epoch, $r_d$, is treated as an independent free parameter and is constrained simultaneously with the cosmological parameters during the statistical analysis. Since the proposed exponential $f(T)$ model is intended to describe the late-time expansion history of the universe, the analysis is performed using the published DESI DR2 BAO measurements without explicitly modelling the radiation-dominated early Universe. Owing to their high statistical precision and extended redshift coverage, the DESI DR2 BAO observations provide stringent constraints on the cosmological parameters of the proposed model.
\subsection{Pantheon Plus \& Pantheon+SH0ES}
The Pantheon Plus compilation is an updated and significantly improved version of the original Pantheon sample, consisting of 1701 light curves from 1550 Type Ia supernovae collected across 18 surveys over the redshift range $0.001 \leq z \leq 2.2613$. Compared with previous compilations, it provides improved photometric calibration, enhanced low-redshift coverage, and a comprehensive treatment of both statistical and systematic uncertainties. The Pantheon+SH0ES sample further includes 77 Cepheid-calibrated Type Ia supernovae, providing an absolute calibration of the distance scale and enabling a direct determination of the Hubble constant $H_0$. Consequently, the combined dataset yields tighter constraints on cosmological parameters by linking relative supernova distances with absolute distance anchors \cite{Scolnic:2018}.

The cosmological parameters are constrained by minimizing the chi-square statistic
\begin{equation}
	\chi_{\rm SN}^{2}
	=
	\Delta \mathbf{D}^{T}
	C_{\rm tot}^{-1}
	\Delta \mathbf{D},
\end{equation}
where $\Delta \mathbf{D}$ denotes the vector of residuals between the observed and theoretical distance moduli, and $C_{\rm tot}$ is the total covariance matrix including both statistical and systematic uncertainties. The theoretical distance modulus is computed from the luminosity distance predicted by the proposed cosmological model. For the Pantheon+SH0ES sample, the Cepheid calibration is incorporated through the corresponding covariance matrix, allowing both relative supernova distances and absolute distance calibration to be treated consistently within a unified likelihood analysis.
\subsection{Observational Gravitational-Wave Catalogs (LIGO--Virgo--KAGRA)}

Gravitational-wave (GW) observations provide an independent probe of the expansion history of the Universe through standard sirens, enabling direct measurements of the luminosity distance without relying on the traditional cosmic distance ladder \cite{Abbott:2016}. In the present work, we employ a publicly available compilation of 90 gravitational-wave events obtained from the Gravitational Wave Open Science Center (GWOSC), including compact binary coalescence events reported by the LIGO--Virgo--KAGRA (LVK) \cite{Acernese:2015,Akutsu:2021} Collaboration in the GWTC catalogs and subsequent public releases \cite{Abbott:2019,Abbott:2021, Abbott:2023,Abbott:2024}. The dataset consists of binary black hole and binary neutron star mergers, for which the observed luminosity distance and the corresponding redshift are available. These measurements are used to constrain the cosmological parameters of the proposed exponential $f(T)$ model.

The gravitational-wave likelihood is constructed using the standard chi-square statistic,
\begin{equation}
	\chi_{\rm GW}^{2}
	=
	\sum_{i=1}^{N}
	\left(
	\frac{d_{L,i}^{\rm obs}-d_{L}^{\rm th}(z_i,\mathbf{\Theta})}
	{\sigma_{d_{L,i}}}
	\right)^2,
\end{equation}
where $d_{L,i}^{\rm obs}$ and $\sigma_{d_{L,i}}$ denote the observed luminosity distance and its corresponding uncertainty for the $i$th gravitational-wave event, respectively, while $d_{L}^{\rm th}(z_i,\mathbf{\Theta})$ represents the theoretical luminosity distance predicted by the proposed exponential $f(T)$ cosmological model. The cosmological parameters are estimated by minimizing the above chi-square statistic.
\section{Results and Discussion}
\subsection{Observational Constraints and Statistical Analysis}
\begin{figure*}[!t]
	
	\centering
	
	\includegraphics[width=0.8\textwidth]{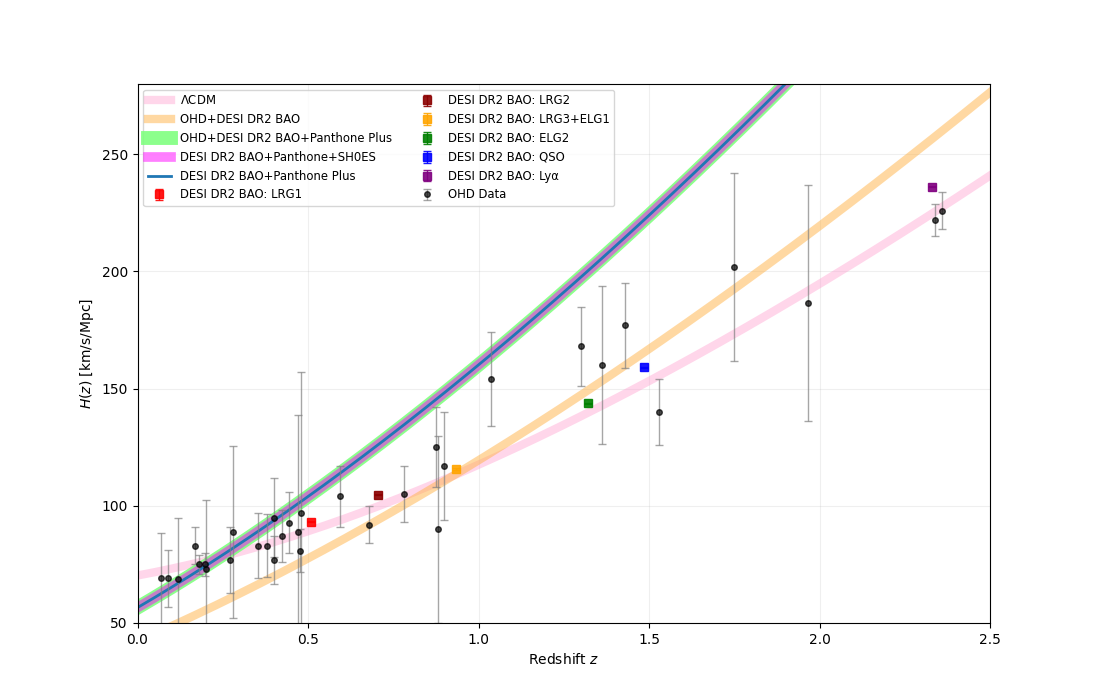}
	
	\caption{Hubble parameter $H(z)$ versus redshift for the $\Lambda$CDM model and the  proposed cosmological model constrained using OHD, DESI DR2 BAO, Pantheon Plus, and SH0ES observational datasets. The OHD measurements with error bars and DESI DR2 BAO data points are displayed, while the shaded regions represent the corresponding model uncertainties.}
	
	\label{fig:cc}
	
\end{figure*}

\begin{figure*}[!t]
	\centering
	\includegraphics[width=0.8\textwidth]{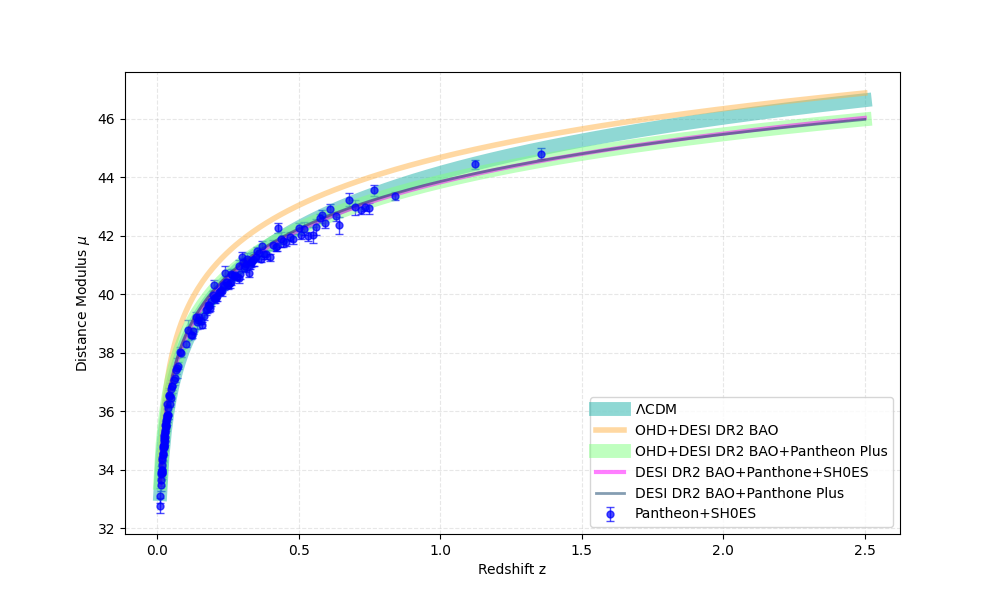}
	\caption{Distance modulus $\mu(z)$ as a function of redshift for $\Lambda$CDM and the alternative cosmological model, constrained by OHD, DESI DR2 BAO, Pantheon Plus, and SH0ES data combinations. Pantheon+SH0ES observations with error bars are shown together with the corresponding best-fit curves.}
	\label{fig:sne}
\end{figure*}

\begin{figure*}[!t]
	\centering
	\includegraphics[width=0.8\textwidth]{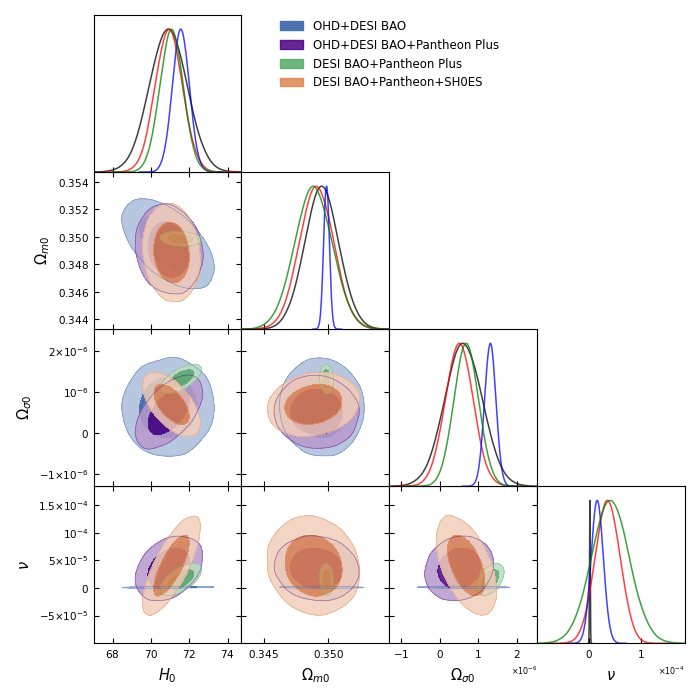}
	\caption{Triangle plot showing the marginalized posterior distributions and joint confidence contours for the cosmological parameters $(H_0, \Omega_{m0}, \Omega_{\sigma0}, \nu)$ of the present cosmological model, constrained using different combinations of OHD, DESI DR2 BAO, Pantheon Plus, and SH0ES datasets. The contours correspond to $1\sigma$ and $2\sigma$ confidence levels.}
	\label{fig:corner}
\end{figure*}

\begin{figure*}[t]
	\centering
	\includegraphics[width=1.0\textwidth]{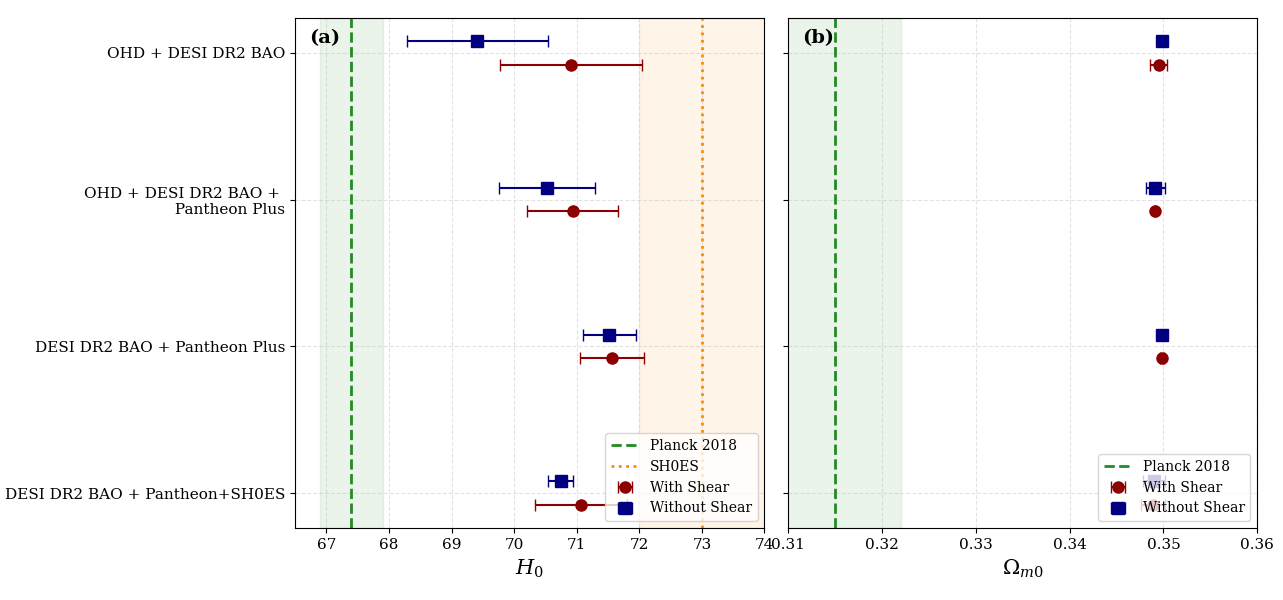}
\caption{Forest plots showing the observational constraints on the cosmological parameters $H_0$ and $\Omega_{m0}$ obtained from different combinations of datasets. The shaded vertical bands denote the Planck 2018 and SH0ES reference regions.}
	\label{fig:h0}
\end{figure*}

In this subsection, we examine the consistency of the present anisotropic exponential $f(T)$ cosmological model with observations using different combinations of OHD, DESI DR2 BAO, Pantheon Plus, and SH0ES data. The Hubble expansion history and distance modulus are reconstructed, and the model parameters are estimated using statistical techniques. We reconstruct the Hubble parameter and the distance modulus and constrain cosmological parameters using statistical methods. In addition, the effects of shear contributions on the cosmological evolution are examined through comparisons with the standard $\Lambda$CDM scenario. Figures~\ref{fig:cc}--\ref{fig:h0} and Tables~\ref{tab:chi}--\ref{tab:bestfit} summarize the observational constraints, statistical analysis, and parameter estimation results obtained for the present cosmological framework.

In Fig.~\ref{fig:cc}, we illustrates the evolution of the Hubble parameter $H(z)$ as a function of redshift. The black points with error bars represent the 33 observational Hubble data (OHD) measurements, while the coloured square markers correspond to DESI DR2 BAO measurements at different redshifts. The solid curve denotes the best-fit theoretical model obtained through the least-squares analysis. The solid curves indicate the best-fit results obtained from different dataset combinations, including OHD+DESI DR2 BAO, OHD+DESI DR2 BAO+Pantheon Plus, and OHD+DESI DR2 BAO+Pantheon+SH0ES. The $\Lambda$CDM model is also shown for comparision. The shaded regions represent the corresponding confidence intervals. The inclusion of the Pantheon+SH0ES dataset improves the constraints at low redshift, leading to a slightly higher normalization of $H(z)$, while maintaining overall consistency with the observational data across the full redshift range.To further evaluate the model performance, we employ the Akaike Information Criterion (AIC) and the Bayesian Information Criterion (BIC), defined as
\[ \text{AIC} = \chi^{2}_{\text{min}} + 2k, \quad 
\text{BIC} = \chi^{2}_{\text{min}} + k \ln(N), \]
where $N$ denotes the number of data points and $k$ is the number of model parameters.
\begin{table*}[t]
	\centering
	
	\caption{Comparison of statistical parameters for the models
		without shear, with shear, and the $\Lambda$CDM model.}
	
	\label{tab:chi}
	
	\scriptsize
	\renewcommand{\arraystretch}{1.2}
	\setlength{\tabcolsep}{4pt}
	
	\begin{ruledtabular}
		
		\begin{tabular}{lcccccccccccc}
			
			Dataset
			&
			\multicolumn{4}{c}{Without Shear}
			&
			\multicolumn{4}{c}{With Shear}
			&
			\multicolumn{4}{c}{$\Lambda$CDM}
			\\
			
			\cline{2-5}
			\cline{6-9}
			\cline{10-13}
			
			& $\chi^2_{\rm min}$ & $\chi^2_{\rm red}$
			& AIC &	BIC
				& $\chi^2_{\rm min}$ & $\chi^2_{\rm red}$
			& AIC &	BIC
				& $\chi^2_{\rm min}$ & $\chi^2_{\rm red}$
			& AIC &	BIC
			\\
			
			\hline
			
			OHD + DESI DR2 BAO
			& 32.10	& 0.747	& 38.10	& 43.59
			& 32.10	& 0.764
			&
			40.10
			&
			47.42
			&
			32.10
			&
			0.730
			&
			36.10
			&
			39.76
			\\
			
			OHD + DESI DR2 BAO
			+\\ Pantheon Plus
			&
			1559.97
			&
			0.961
			&
			1565.97
			&
			1582.15
			&
			1545.25
			&
			0.953
			&
			1553.25
			&
			1574.82
			&
			1559.97
			&
			0.961
			&
			1563.97
			&
			1574.76
			\\
			
			DESI DR2 BAO
			+\\ Pantheon + SH0ES
			&
			1555.94
			&
			0.972
			&
			1561.94
			&
			1578.08
			&
			1536.12
			&
			0.961
			&
			1544.12
			&
			1565.64
			&
			1555.94
			&
			0.972
			&
			1559.94
			&
			1570.70
			\\
			
			DESI DR2 BAO
			+\\ Pantheon Plus
			&
			1540.58
			&
			0.969
			&
			1546.58
			&
			1562.70
			&
			1520.74
			&
			0.957
			&
			1528.74
			&
			1550.23
			&
			1540.58
			&
			0.968
			&
			1544.58
			&
			1555.33
			\\
			
		\end{tabular}
		
	\end{ruledtabular}
	
\end{table*}
\begin{table*}[!t]
	\centering
	
	\caption{Differences in information criteria for the models with and without shear relative to the $\Lambda$CDM model, along with the direct effect of shear.}
	
	\label{tab:values}
	
	\scriptsize
	\renewcommand{\arraystretch}{2.6}
	\setlength{\tabcolsep}{8pt}

	\begin{ruledtabular}
		
		\begin{tabular}{lcccccc}
			
			& \multicolumn{2}{c}{\makecell{Without Shear  vs $\Lambda$CDM}}
			& \multicolumn{2}{c}{\makecell{With Shear  vs $\Lambda$CDM}}
			& \multicolumn{2}{c}{Shear Effect} \\
			
			\colrule
			
			\textbf{Dataset}
			& $|\Delta\mathrm{AIC}|$
			 & $|\Delta\mathrm{BIC}|$
			& $|\Delta\mathrm{AIC}|$ & $|\Delta\mathrm{BIC}|$
			& $|\Delta\mathrm{AIC}|$ & $|\Delta\mathrm{BIC}|$ \\
			
			\colrule
			
			\makecell[l]{OHD + DESI DR2 BAO}
			
			& 2.00 & 3.83
			& 4.00 & 7.66
			& 2.00 & 3.83 \\
			
			\makecell[l]{OHD + DESI DR2 BAO \\ + Pantheon Plus}
			
			& 2.00 & 7.39
			& 10.72 & 0.06
			& 12.72 & 7.33 \\
			
			\makecell[l]{DESI DR2 BAO + \\ Pantheon + SH0ES}
			
			& 2.00 & 7.38
			& 15.82 & 5.06
			& 17.82 & 12.44 \\
			
			\makecell[l]{DESI DR2 BAO + \\ Pantheon Plus}
			
			& 2.00 & 7.37
			& 15.84 & 5.10
			& 17.84 & 12.47 \\
			
		\end{tabular}
		
	\end{ruledtabular}
	
\end{table*}
Tables~\ref{tab:chi} and~\ref{tab:values} summarise the statistical analysis for the models without shear, with shear, and the standard $\Lambda$CDM model using different combinations of observational datasets. Table~\ref{tab:chi} presents the reduced chi-square values along with the corresponding AIC and BIC statistics. All models provide good fits to the data, as indicated by reduced chi-square values close to unity. Comparing AIC and BIC values shows that the statistical preference depends on the combination of observational datasets. For the OHD+DESI DR2 BAO combination, the $\Lambda$CDM model yields the lowest AIC and BIC values, while the shear-inclusive model provides lower AIC and BIC values for the combined datasets.  

Table~\ref{tab:values} shows the differences in information criteria relative to the $\Lambda$CDM model, together with the direct effect of including shear. For the model without shear, the values of $\Delta \mathrm{AIC}\approx2$ across all dataset combinations indicate a close statistical comparison with the $\Lambda$CDM scenario, while the relatively larger $\Delta\mathrm{BIC}$ values reflect the stronger penalty imposed by BIC for the additional model parameter. For the model with shear, the information-criterion differences depend on the dataset combination, with the combined datasets generally showing an improved statistical fit compared with the corresponding non-shear case. The direct effect of shear becomes more pronounced for the combined datasets, where the inclusion of shear leads to lower $\chi^2$ and AIC values, indicating that the shear contribution can improve the statistical fit under the present observational constraints.

 In  Fig.~\ref{fig:sne}, we describes the evolution of the distance modulus $\mu(z)$ as a function of redshift for different dataset combinations. The blue data points with error bars represent the Pantheon+SH0ES supernova observations, while the solid curves correspond to the best-fit results obtained from the OHD+DESI DR2 BAO, OHD+DESI DR2 BAO+Pantheon Plus, DESI DR2 BAO+Pantheon Plus, and DESI DR2 BAO+Pantheon+SH0ES datasets. The $\Lambda$CDM prediction is also displayed for comparison.
It is evident that all models exhibit good agreement with the supernova observations throughout the considered redshift range. The incorporation of Pantheon Plus and SH0ES data provides tighter constraints on the cosmological evolution, particularly in the low-redshift region. Moreover, only small deviations are observed among the different dataset combinations, suggesting that the reconstructed cosmological behaviour remains broadly consistent with the observational supernova data.

 The cosmological parameters are constrained within a Bayesian inference framework by constructing the total chi-square function as the sum of the individual chi-square contributions from the considered datasets. Parameter estimation is performed using the Markov Chain Monte Carlo (MCMC) sampling technique with $60{,}000$ samples, from which the best-fit values and the corresponding covariance matrix are obtained. Uniform priors are imposed on all free parameters, as summarized in Table~\ref{tab:params}. The corresponding best-fit values for the models with and without shear contributions obtained from different observational datasets are presented in Table~\ref{tab:bestfit}.

  In Fig.~\ref{fig:corner}, we presents the joint and marginalized posterior distributions of the model parameters $H_{0}$, $\Omega_{m0}$, $\Omega_{\sigma0}$, and $\nu$ obtained from different observational dataset combinations, namely OHD+DESI DR2 BAO, OHD+DESI DR2 BAO+Pantheon Plus, DESI DR2 BAO+Pantheon Plus, and DESI DR2 BAO+Pantheon+SH0ES. The diagonal panels represent the one-dimensional marginalized distributions of each parameter, while the off-diagonal panels show the corresponding two-dimensional confidence contours at the $1\sigma$ and $2\sigma$ confidence levels. 

 It is observed that the Hubble parameter $H_{0}$ is constrained within the approximate range $70$--$72~\mathrm{km,s^{-1}, Mpc^{-1}}$ for all dataset combinations, with tighter constraints obtained when Pantheon Plus and SH0ES data are included. The matter density parameter $\Omega_{m0}$ remains close to $\Omega_{m0}\approx0.35$, indicating consistency with the concordance cosmological scenario. The anisotropic density parameter $\Omega_{\sigma0}$ is found to be very small, suggesting that the universe approaches isotropy at late times. Moreover,the model parameter $\nu$ remain constrained,
  with $\nu$ taking small values of order $10^{-5}$--$10^{-4}$, implying that the exponential correction contributes only as a small perturbative effect in the teleparallel gravity framework. Overall, the contour regions corresponding to different datasets show substantial overlap, demonstrating that the proposed exponential $f(T)$ model remains compatible with the current observational data.  
 
\begin{table}[!t]
	\centering
	
	\caption{Uniform prior ranges assumed for the cosmological and exponential $f(T)$ model parameters in the MCMC observational analysis.}
	
	\label{tab:params}
	
	\small
	\renewcommand{\arraystretch}{1.1}
	
	\begin{ruledtabular}
		
		\begin{tabular}{ccc}
			
			\textbf{Parameter} & \textbf{Prior Range} & \textbf{Interpretation} \\
			
			\colrule
			
			$H_0$ & $(60,100)$ & Hubble constant \\
			
			$\Omega_{m0}$ & $(0.2,0.4)$ & Matter density parameter \\
			
			$\Omega_{\sigma0}$ & $(10^{-9},10^{-3})$ & Shear density parameter \\
			
			
			$\nu$ & $(10^{-12},10^{-6})$ & Exponential decay parameter \\
			
			$M$ & $(-21,-18)$ & Absolute magnitude parameter \\
			
			$r_d$ & $(130,180)$ & Sound horizon scale \\
			
		\end{tabular}
		
	\end{ruledtabular}
	
\end{table}
\begin{table*}[!t]
	\centering
	
	\caption{Comparison of cosmological parameters for the $f(T)$ cosmological model without shear and with shear contributions using different observational datasets.}
	
	\label{tab:bestfit}
	
	\scriptsize
	\renewcommand{\arraystretch}{2.6}
	\setlength{\tabcolsep}{8pt}

	\begin{ruledtabular}
		
		\begin{tabular}{lcc}
			
			Dataset & Without Shear & With Shear \\
			
			\colrule
			
			OHD + DESI DR2 BAO
			
			&
			
			\makecell[c]{
				$H_0=69.41^{+1.59}_{-0.65}$\\
				$\Omega_{m0}=0.3498^{+0.0001}_{-0.0005}$\\
				$q_0=-0.474  $\\
				$w_0^\mathrm{eff}=-0.649  $\\
			}
			
			&
			
			\makecell[c]{
				$H_0=70.90^{+1.00}_{-1.26}$\\
				$\Omega_{m0}=0.3491^{+0.0004}_{-0.0001}$\\
				$\Omega_{\sigma0}=(6.20^{+5.21}_{-4.53})\times10^{-7}$\\
				$q_0=-0.463$\\
				$w_0^\mathrm{eff}=-0.642$\\
			}
			
			\\
			
			\makecell[l]{OHD + DESI DR2 BAO\\+ Pantheon Plus}
			
			&
			
			\makecell[c]{
				$H_0=70.74^{+0.17}_{-0.24}$\\
				$\Omega_{m0}=0.3490^{+0.0007}_{-0.0016}$\\
				$q_0=-0.126   $\\
				$w_0^\mathrm{eff}=-0.417   $\\
			}
			
			&
			
			\makecell[c]{
				$H_0=71.07^{+0.75}_{-0.73}$\\
				$\Omega_{m0}=0.3489^{+0.0009}_{-0.0017}$\\
				$\Omega_{\sigma0}=(6.90^{+2.85}_{-4.97})\times10^{-7}$\\
				$q_0=-0.212$\\
				$w_0^\mathrm{eff}=-0.474$\\
			}
			
			\\
			\makecell[l]{DESI DR2 BAO\\+ Pantheon Plus}
			
			&
			
			\makecell[c]{
				$H_0=70.53^{+0.49}_{-1.04}$\\
				$\Omega_{m0}=0.3491^{+0.0006}_{-0.0013}$\\
				$q_0=0.104$\\
				$w_0^\mathrm{eff}=-0.264   $\\
			}
			
			&
			
			\makecell[c]{
				$H_0=70.93^{+0.93}_{-0.52}$\\
				$\Omega_{m0}=0.3491^{+0.0006}_{-0.0016}$\\
				$\Omega_{\sigma0}=(5.06^{+4.40}_{-3.67})\times10^{-7}$\\
				$q_0=-0.337$\\
				$w_0^\mathrm{eff}=-0.558$\\
			}
			
			\\

			\makecell[l]{DESI DR2 BAO\\+ Pantheon + SH0ES}
			
			&
			
			\makecell[c]{
				$H_0=71.52^{+0.43}_{-0.42}$\\
				$\Omega_{m0}=0.3499^{+0.0001}_{-0.0002}$\\
				$q_0=-0.097    $\\
				$w_0^\mathrm{eff}=-0.398   $\\
			}
			
			&
			
			\makecell[c]{
				$H_0=71.57^{+0.46}_{-0.55}$\\
				$\Omega_{m0}=0.3499^{+0.0001}_{-0.0003}$\\
				$\Omega_{\sigma0}=(1.31^{+0.12}_{-0.18})\times10^{-6}$\\
				$q_0=-0.183$\\
				$w_0^\mathrm{eff}=-0.455$\\
			}
			
			\\

		\end{tabular}
		
	\end{ruledtabular}
\end{table*}

 In Fig.~\ref{fig:h0}, we compares the parameter constraints obtained in the presence and absence of shear contributions for various observational dataset combinations. In panel (a), the reconstructed values of the Hubble parameter $H_0$ are found to lie between the Planck 2018 and SH0ES measurements, indicating consistency with late--time cosmological observations. The inclusion of shear produces a mild shift toward slightly larger values of $H_0$ for certain datasets. Panel (b) presents the corresponding constraints on the matter density parameter $\Omega_{m0}$, where most estimates remain compatible with the Planck 2018 limits within the uncertainty range. The comparatively larger error bars for the DESI DR2 BAO + Pantheon Plus combination indicate weaker parameter constraints due to the reduced observational coverage. Overall, the results demonstrate that the proposed model remains observationally viable for both shear and non--shear scenarios.
\subsection{BAO and Gravitational Wave Analysis}
\begin{figure*}[t]
	\centering
	\includegraphics[width=0.8\textwidth]{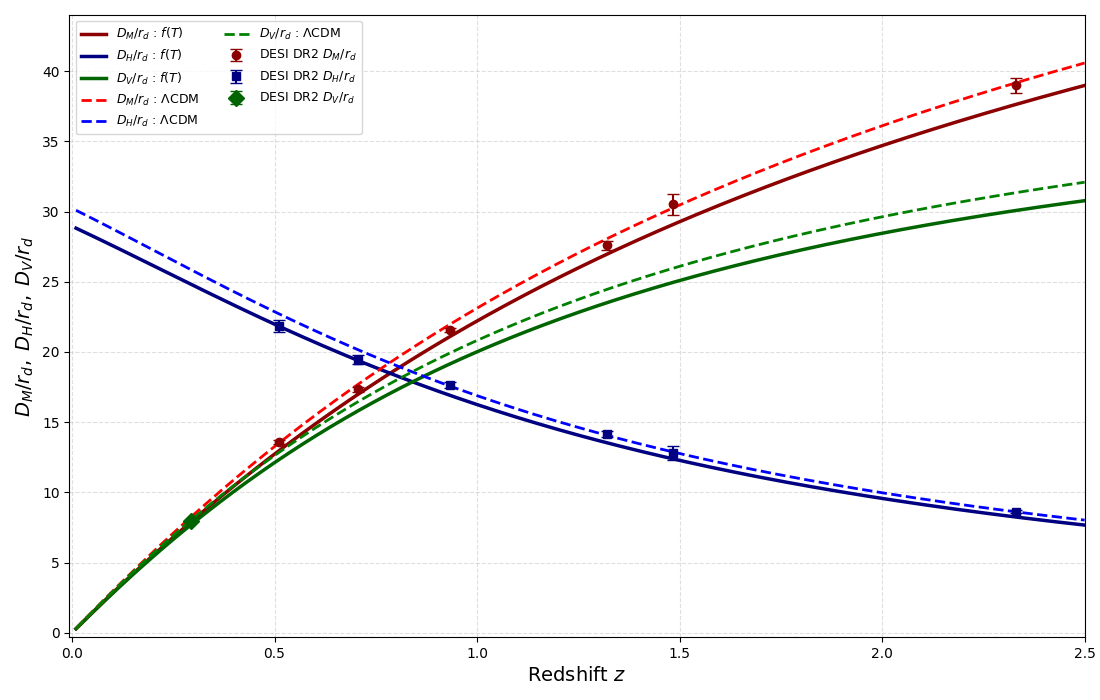}
	\caption{Comparison of the reconstructed BAO observables $D_M/r_d$, $D_H/r_d$, and $D_V/r_d$ with the DESI DR2 measurements. The solid lines show the best--fit evolution of the proposed cosmological model, whereas the dashed lines indicate the corresponding predictions of the $\Lambda$CDM model.}
	\label{fig:bao}
\end{figure*}

\begin{figure*}[t]
	\centering
	\includegraphics[width=1.0\textwidth]{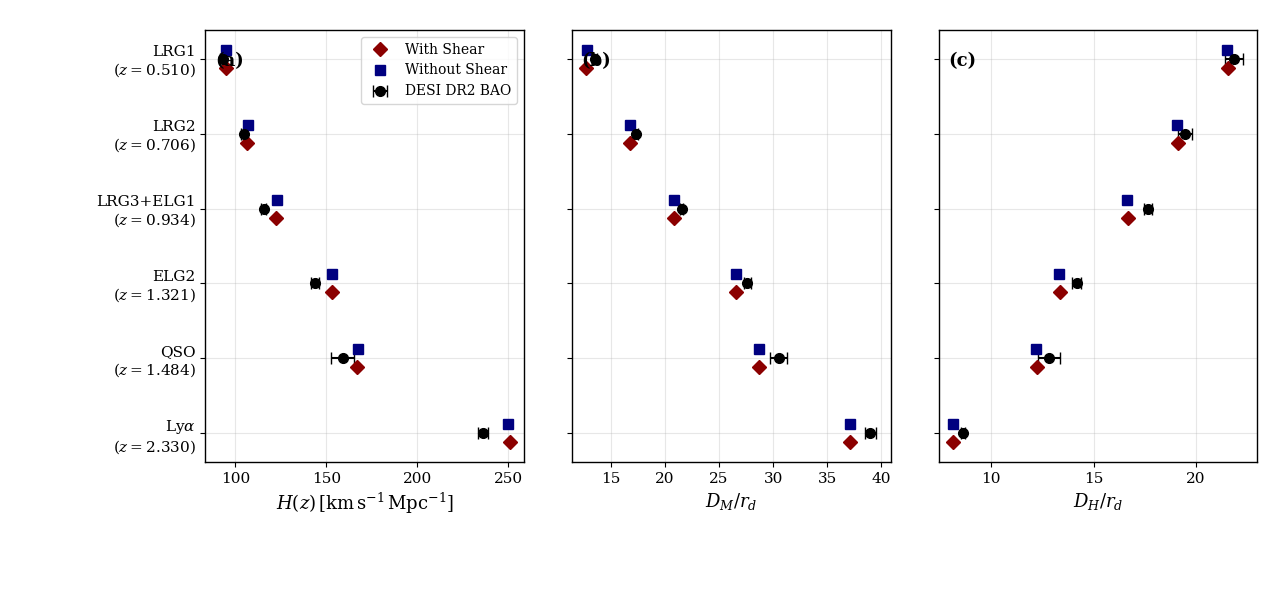}
\caption{
	Forest-plot comparison of the reconstructed cosmological quantities
	$H(z)$, $D_M/r_d$, and $D_H/r_d$ with the DESI DR2 BAO measurements at
	different redshifts. The red diamond markers represent the best-fit
	predictions of the model including shear effects, while the blue square
	markers correspond to the non-shear case. The black points with error
	bars denote the DESI DR2 observational data. Panels (a), (b), and (c)
	show the comparisons for the Hubble parameter, transverse comoving
	distance, and radial BAO distance scale, respectively.
}
	\label{fig:three}
\end{figure*}

\begin{figure*}[t]
	\centering
	\includegraphics[width=0.8\textwidth]{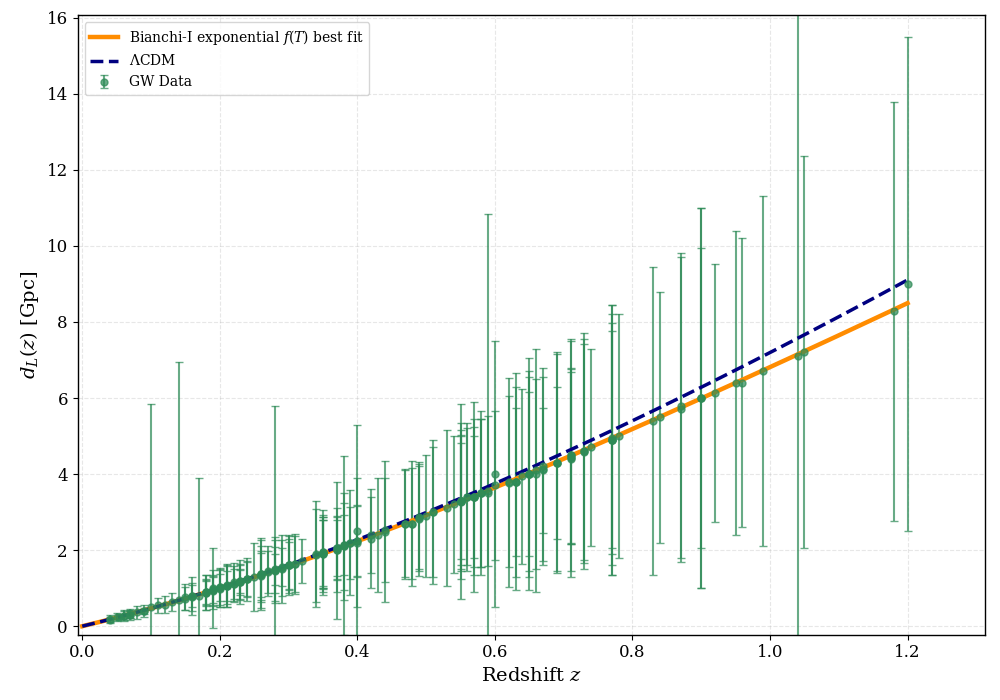}
\caption{Evolution of the luminosity distance $d_L(z)$ as a function of redshift for the reconstructed cosmological model together with the gravitational wave observational dataset. The solid curve represents the best--fit prediction of the proposed framework, while the dashed curve corresponds to the standard $\Lambda$CDM cosmology. The green data points with error bars denote the observed gravitational wave measurements.}
	\label{fig:gw}
\end{figure*}

 In this subsection, we examine the observational consistency of the proposed cosmological model using baryon acoustic oscillation (BAO) and gravitational wave datasets. The reconstructed distance measures and luminosity distance evolution are analysed through comparisons with DESI DR2 BAO observations and gravitational wave standard siren measurements. The corresponding results also allow us to investigate the influence of shear contributions and deviations from the standard $\Lambda$CDM cosmology.
 Figures~\ref{fig:bao}--\ref{fig:gw} summarize the reconstructed BAO distance measures, DESI DR2 BAO comparisons, and gravitational wave constraints obtained for the present cosmological framework.

 In Fig.~\ref{fig:bao}, we depict the evolution of the BAO distance measures $D_M/r_d$, $D_H/r_d$, and $D_V/r_d$ as functions of redshift for the proposed cosmological model using the DESI DR2 BAO dataset. The reconstructed curves are in good agreement with the DESI DR2 measurements over the considered redshift range. The quantity $D_M/r_d$ increases with redshift, whereas $D_H/r_d$ decreases, consistent with the expected cosmic expansion history. Similarly, $D_V/r_d$ exhibits a smooth monotonic growth and closely follows the observational data. Overall, the small deviation from the corresponding $\Lambda$CDM predictions indicates that the proposed model remains compatible with the current DESI DR2 BAO observations.
 
 In Fig.~\ref{fig:three}, we presents a comparative analysis between the theoretical predictions of the proposed cosmological model and the DESI DR2 BAO observations over the considered redshift interval. The reconstructed values of $H(z)$ displayed in panel (a) remain close to the observational measurements, indicating that both shear and non--shear configurations successfully reproduce the late--time expansion history. In panel (b), the transverse distance ratio $D_M/r_d$ shows only minor deviations from the DESI measurements, with the discrepancies remaining within the observational uncertainty limits. Similarly, panel (c) demonstrates that the predicted radial distance scale $D_H/r_d$ follows the observed trend across the full redshift range. A mild separation between the shear and non--shear cases can be noticed at higher redshifts, reflecting the influence of anisotropic contributions on the cosmic dynamics. Nevertheless, the overall agreement with the DESI DR2 BAO data confirms the observational consistency and stability of the proposed framework.

 In Fig.~\ref{fig:gw}, we illustrates the behaviour of the luminosity distance over the considered redshift interval using gravitational wave observations. The reconstructed trajectory obtained from the proposed model follows the overall trend of the observational data and remains close to the standard cosmological prediction throughout the analysed region. At lower redshifts, the theoretical curves nearly overlap, indicating minimal deviation from the conventional expansion scenario. A small separation between the two cosmological evolutions becomes noticeable toward higher redshift values, reflecting the influence of the additional model contributions on the late--time dynamics. The comparatively larger uncertainties associated with several gravitational wave events are consistent with the present observational limitations of standard siren measurements. Nevertheless, the agreement between the reconstructed curve and the dataset demonstrates that the model successfully reproduces the observed luminosity distance behaviour within the current error bounds.
\subsection{Expansion History and Dynamical Analysis}
\begin{figure*}[t]
	\centering
	\includegraphics[width=0.8\textwidth]{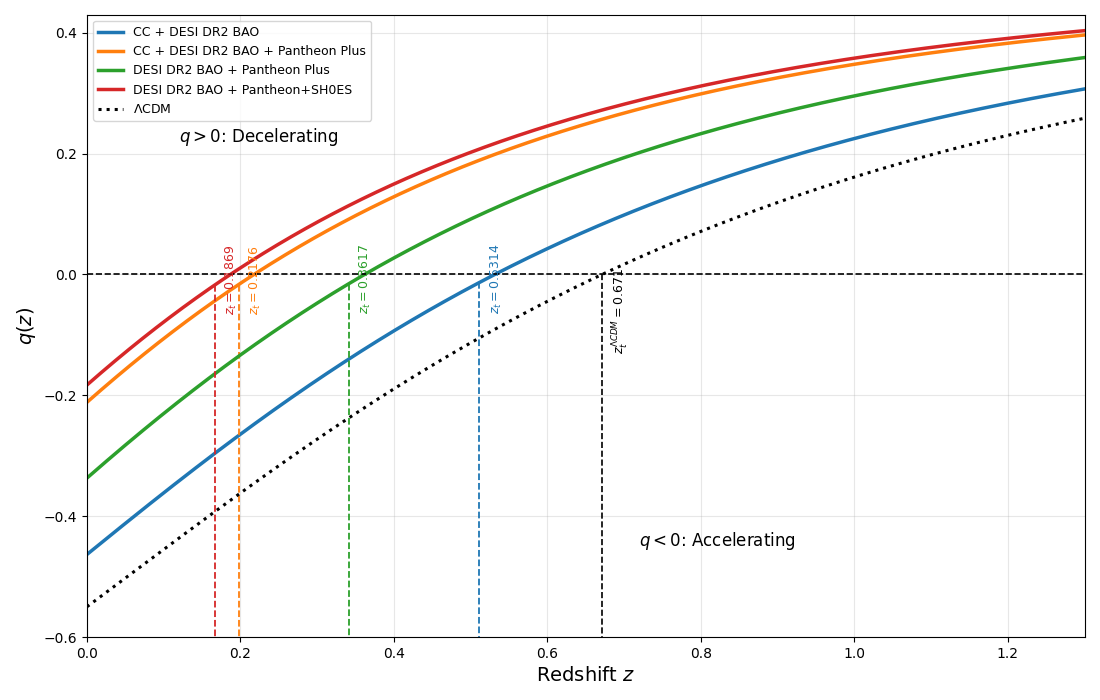}
	\caption{Redshift evolution of the deceleration parameter for the reconstructed cosmological model using different observational dataset combinations. The dashed curve corresponds to the standard $\Lambda$CDM cosmology, while the shaded regions distinguish the accelerating and decelerating expansion regimes.}
	\label{fig:dec}
\end{figure*}

\begin{figure*}[t]
	\centering
	\includegraphics[width=0.8\textwidth]{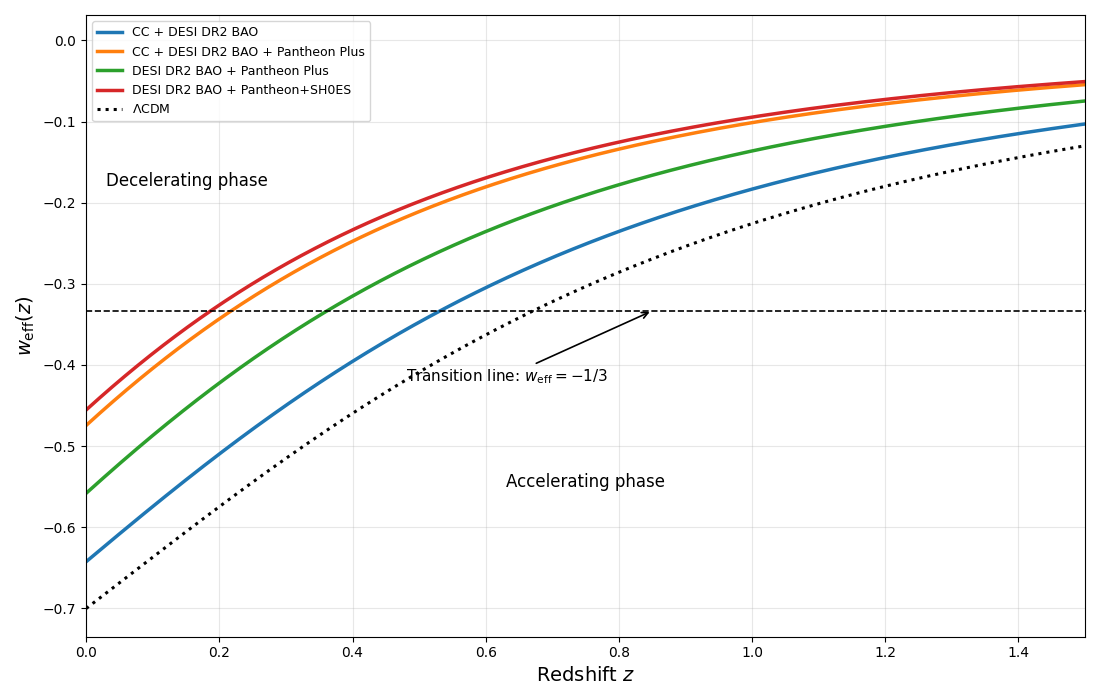}
	\caption{Effective equation of state parameter $w^{\mathrm{eff}}(z)$ as a function of redshift for $\Lambda$CDM and the anisotropic $f(T)$ gravity model, constrained by OHD, DESI DR2 BAO, Pantheon Plus, and SH0ES datasets. The horizontal line at $w^{\mathrm{eff}}=-1/3$ marks the transition between decelerating and accelerating expansion, with shaded regions indicating the corresponding regimes.}
	\label{fig:eos} \end{figure*}

In this subsection, we investigate the expansion history and dynamical behaviour of the proposed anisotropic cosmological model using the deceleration parameter and the effective equation of state parameter. These diagnostics provide insight into the evolution of the cosmic expansion and the transition from a matter-dominated decelerating phase to the present accelerated epoch. The dynamical properties of the model are further examined through the evolution of these parameters, while comparisons with the standard $\Lambda$CDM cosmology are performed to assess the consistency and viability of the reconstructed cosmological scenario.
Figures~\ref{fig:dec} and \ref{fig:eos} summarize the geometrical diagnostics and dynamical evolution of the proposed cosmological model obtained from different observational dataset combinations.

\textbf{Deceleration parameter $q(z)$:}
 The deceleration parameter $q(z)$ is an important geometrical quantity used to describe the expansion dynamics of the universe. It is defined as
\begin{equation}
	q(z)=-1+\frac{(1+z)}{H(z)}\frac{dH(z)}{dz}.
\end{equation}

 A positive value of $q(z)$ represents a decelerating universe, while a negative value indicates an accelerating phase of cosmic expansion. Hence, the behaviour of the deceleration parameter provides useful information about the transition from the matter--dominated era to the present accelerated epoch.
 In Fig.~\ref{fig:dec}, we demonstrates the dynamical behaviour of the cosmic expansion through the evolution of the deceleration parameter $q(z)$. The reconstructed curves show that the universe evolves from a positive deceleration regime at higher redshifts toward negative values at later times, indicating the onset of accelerated expansion. The crossing point near $q(z)=0$ represents the cosmic transition epoch between these two phases. Minor differences among the trajectories arise from the distinct observational datasets employed in the analysis, although the overall behaviour remains consistent throughout the considered redshift interval. The close agreement with the reference cosmological scenario indicates that the proposed anisotropic framework can successfully reproduce the observed late--time acceleration of the universe.
 
 \textbf{Effective Equation of State Parameter}($w^\mathrm{eff}(z)$):
 The effective equation of state parameter $w^{\mathrm{eff}}(z)$ provides a useful description of the overall expansion behaviour of the universe by relating the effective pressure and energy density of the cosmic contents. In terms of the Hubble parameter, it can be expressed as
\begin{equation}
	w^{\mathrm{eff}}(z)=-1+\frac{(1+z)}{3H(z)}\frac{dH(z)}{dz}.
\end{equation}

 The evolution of $w^{\mathrm{eff}}(z)$ helps to distinguish different cosmological phases during the history of the universe. Values satisfying $w^{\mathrm{eff}}<-1/3$ indicate accelerated expansion, whereas $w^{\mathrm{eff}}=0$ and $w^{\mathrm{eff}}=1/3$ correspond to matter and radiation dominated epochs, respectively.
 In Fig.~\ref{fig:eos}, we describe the evolution of the effective equation of state parameter ${w^\mathrm{eff}}(z)$ for different observational dataset combinations. The reconstructed trajectories lie in the quintessence region at low redshifts and gradually approach less negative values at higher redshifts. The crossing near the transition boundary $w^{\mathrm{eff}}=-1/3$ distinguishes the change between accelerated and decelerated cosmic expansion. Small variations among the curves arise due to the influence of different observational constraints, although the overall behaviour remains consistent with the standard cosmological scenario. The present negative values of the effective equation of state parameter indicate that the proposed anisotropic framework successfully reproduces the observed late--time acceleration of the universe without entering the phantom regime.  
 
\subsection{Dynamical System Analysis}

The dynamical behaviour of the proposed anisotropic exponential $f(T)$ model is examined using the effective cosmological evolution obtained in the regime $|\nu T|\ll1$. For this purpose, we define the dimensionless variables

\begin{equation}
\chi=\frac{\Omega_{m0}a^{-3}}{(1+\mu \nu)E^2},
\qquad
\psi=\frac{\Omega_{\sigma0}a^{-6}}{E^2},
\qquad
\lambda=-\frac{\mu}{6{H_0^{2}}(1+\mu \nu){E^2}},
\end{equation}

where $	E=\frac{H}{H_0}$.
The evolution equations can be written in the autonomous form

\begin{equation}
	\chi' = 3\chi(\chi+2\psi-1),
	\qquad
	\psi' = 3\psi(\chi+2\psi-2).
\end{equation}
Derivatives are taken with respect to  $N=\ln a$.
Using the background solution given by Eq.~\eqref{20}, we obtain
\begin{equation}
	w^{\rm eff}
	=
	\chi +2\psi -1.
\end{equation}

The equilibrium points are obtained from
$	\chi^{'}=0,
\psi^{'}=0$ and  their stability is determined from the Jacobian matrix

\begin{equation}
	J=
	\begin{pmatrix}
		6\chi+6\psi-3 & 6\chi \\
		3\psi & 3\chi+12\psi-6
	\end{pmatrix}.
\end{equation}

\paragraph{Shear-dominated point:}
The equilibrium point $P_{1}=(0,1)$ corresponds to a shear-dominated era with $w^{\rm eff}=1$. The associated eigenvalues are $(\lambda_{1},\lambda_{2})=(3,6)$, showing that the point is an unstable node. Consequently, this phase characterizes the early evolution of the cosmological model.
\paragraph{Matter-dominated point:}
For the equilibrium solution $P_{2}=(1,0)$, one finds $w^{\rm eff}=0$, indicating a matter-dominated epoch. The associated eigenvalues are $(\lambda_{1},\lambda_{2})=(3,-3)$, which imply a saddle point. Such a transient state allows the cosmological evolution to pass through a matter era before approaching the accelerating attractor.

\paragraph{de-Sitter point:}
For the equilibrium solution $P_{3}=(0,0)$, one obtains $w^{\rm eff}=-1$, which characterizes a de-Sitter epoch. The eigenvalues of the Jacobian matrix are $(\lambda_{1},\lambda_{2})=(-3,-6)$, indicating a stable node. Therefore, the model naturally evolves towards a dark-energy-dominated accelerated phase at late times. The properties of the equilibrium points are summarized in Table~\ref{tab:critical}.

\begin{table*}[t]
	\caption{Phase-space fixed points and their cosmological characteristics.}
	\label{tab:critical}
	\centering
		\scriptsize
	\renewcommand{\arraystretch}{2.6}
	\setlength{\tabcolsep}{8pt}
	\begin{tabular}{cccccc}
		\hline\hline
		Fixed Point & $(\chi,\psi)$ & $w_{\rm eff}$ &
		$(\lambda_1,\lambda_2)$ &
		Dynamical Nature &
		Cosmological Description \\
		\hline
		$P_1$ & $(0,1)$ & $1$ &
		$(3,6)$ &
		Repeller &
		Anisotropy-dominated state \\
		$P_2$ & $(1,0)$ & $0$ &
		$(3,-3)$ &
		Saddle &
		Matter-dominated stage \\
		$P_3$ & $(0,0)$ & $-1$ &
		$(-3,-6)$ &
		Attractor &
		Accelerated de Sitter phase \\
		\hline\hline
	\end{tabular}
\end{table*}
The obtained critical points indicate a viable late-time cosmological
evolution. The universe evolves from an anisotropic shear-dominated phase
towards a transient matter-dominated epoch and eventually settles into a
stable de-Sitter accelerated state. Therefore, the present exponential
$f(T)$ model successfully reproduces the sequence

\begin{equation}
	{\rm Shear}
	\rightarrow
	{\rm Matter}
	\rightarrow
	{\rm de\ Sitter},
\end{equation}

which characterizes the late-time cosmological dynamics of the model. 
\subsection{Model Stability and Physical Viability}

The physical viability of the proposed exponential $f(T)$ model can be examined through the stability conditions associated with the derivatives of the functional form 
\begin{equation}
	f(T)=T+\mu e^{\nu T}.
\end{equation}
The corresponding first and second derivatives with respect to the torsion scalar are given by
\begin{equation}
	f_T = 1 + \mu \nu e^{\nu T},
\end{equation}
and 
\begin{equation}
	f_{TT} = \mu \nu^2 e^{\nu T}.
\end{equation}
For a physically acceptable cosmological model, the condition $f_T>0$ must be satisfied to avoid ghost instabilities in the gravitational sector, while a positive value also ensures a positive and finite effective gravitational coupling defined as $G_{\rm eff}=G/f_T$. Additionally, the stability of scalar cosmological perturbations requires the quantity $f_T+2Tf_{TT}$ to remain positive throughout the cosmological evolution. 

Using the observationally constrained best-fit values of $\mu$ and $\nu$, all these critical conditions are satisfied, as shown in Table~\ref{tab:stability}.
 The stability indicators obtained from different observational dataset combinations are summarized in Table~\ref{tab:best1} and Table~\ref{tab:stability}. Furthermore, the model naturally reduces to the teleparallel equivalent of General Relativity in the limit $\mu \rightarrow 0$.

\begin{table*}[ht]
	\centering
	\caption{Comparison of the constrained exponential $f(T)$ model parameters with and without shear contributions for different observational datasets together with the corresponding background stability condition.}
	\label{tab:best1}
	\scriptsize
	\renewcommand{\arraystretch}{2.6}
	\setlength{\tabcolsep}{8pt}
	\begin{ruledtabular}
		\begin{tabular}{lccc}
			
			Dataset & Without Shear & With Shear  \\
			
			\colrule
			
			\makecell[l]{OHD + DESI DR2 BAO}
		&
		\begin{tabular}{c}
			$\mu = 0.524^{+0.970}_{-0.288}$ \\
			$\nu = (2.08^{+5.90}_{-4.15})\times10^{-8}$
		\end{tabular}
		&
		\begin{tabular}{c}
			$\mu = 0.538^{+0.570}_{-0.416}$ \\		
			$\nu = (1.21^{+8.73}_{-5.52})\times10^{-7}$
		\end{tabular}
		
		\\
		
		\makecell[l]{OHD + DESI DR2 BAO \\ + Pantheon Plus}
		&
		\begin{tabular}{c}
			$\mu = 0.549^{+0.337}_{-0.312}$ \\	$\nu = (5.34^{+4.40}_{-3.76})\times10^{-5}$
		\end{tabular}
		&
		\begin{tabular}{c}
			$\mu = 0.261^{+0.295}_{-0.173}$ \\	$\nu = (4.07^{+5.55}_{-2.93})\times10^{-5}$
		\end{tabular}
		
		\\
		
		\makecell[l]{DESI DR2 BAO \\ + Pantheon Plus}
		&
		\begin{tabular}{c}
			$\mu = 0.322^{+0.498}_{-0.226}$ \\	$\nu = (1.41^{+1.64}_{-1.01})\times10^{-4}$
		\end{tabular}
		&
		\begin{tabular}{c}
			$\mu = 0.349^{+0.368}_{-0.248}$ \\	$\nu = (3.53^{+3.09}_{-2.31})\times10^{-5}$
		\end{tabular}
		
		\\
		
		\makecell[l]{DESI DR2 BAO \\ + Pantheon + SH0ES}
		&
		\begin{tabular}{c}
			$\mu = 0.170^{+0.129}_{-0.129}$ \\
			$\nu = (5.90^{+3.79}_{-1.37})\times10^{-5}$
		\end{tabular}
		&
		\begin{tabular}{c}
			$\mu = 0.842^{+0.123}_{-0.326}$ \\	$\nu = (1.54^{+1.64}_{-0.80})\times10^{-5}$
		\end{tabular}
		
			\\
			
		\end{tabular}
	\end{ruledtabular}
\end{table*}

\begin{table*}[ht]
	\centering
	\caption{Numerical evaluation of the ghost-free condition ($f_T$), normalized effective gravitational coupling ($G_{\rm eff}/G$), and scalar perturbation stability ($f_T+2Tf_{TT}$) for the exponential $f(T)$ model.}
	\label{tab:stability}
	\begin{ruledtabular}
			\scriptsize
		\renewcommand{\arraystretch}{2.6}
		\setlength{\tabcolsep}{8pt}
		\begin{tabular}{lcccc}
			Dataset &
			$f_T$ &
			$G_{\rm eff}/G$ &
			$f_T+2Tf_{TT}$ 
			\\
			\colrule
			
			OHD + DESI DR2 BAO
			&
			0.999999371
			&
			1.000000629
			&
			0.999999417
			
			\\
			
			OHD + DESI DR2 BAO + Pantheon Plus
			&
			0.999997773
			&
		    1.000002227
			&
			1.000002046
			
			\\
			
			DESI DR2 BAO + Pantheon Plus
			&
			1.000000032
			&
			0.999999968
			&
			1.000000031
			
			\\
			
			DESI DR2 BAO + Pantheon + SH0ES
			&
			0.999991931
			&
			1.000008069
			&
			0.999999548
			
			\\
			
		\end{tabular}
	\end{ruledtabular}
\end{table*}
It is important to note that many commonly studied $f(T)$ functional forms in the literature, including the Linder exponential model, power--law corrections, and logarithmic modifications, are primarily investigated within the isotropic FLRW background \cite{Linder:2010py,Bamba:2012cp}. In contrast, the present work extends the exponential $f(T)$ framework to an anisotropic Bianchi type-I cosmology, thereby providing a more general description of cosmic evolution and testing its observational viability using recent OHD, DESI DR2 BAO, Pantheon Plus, SH0ES, and gravitational-wave data.

The adopted exponential form is mathematically well-behaved and remains in good agreement with the observational data. For all datasets, $f_T \approx 1$ implies only small deviations from standard gravity, indicating that the model closely follows the successful late-time cosmological evolution of the $\Lambda$CDM scenario. Moreover, the observational constraints favour a small but non-zero anisotropic contribution, consistent with the expected late-time isotropization of the universe.

As shown in Table~\ref{tab:stability}, all datasets yield $f_T+2Tf_{TT}>0$, demonstrating that the model remains stable against small cosmological perturbations within the observationally allowed parameter region. Therefore, the proposed exponential $f(T)$ functional form provides a stable, ghost-free, and physically viable anisotropic cosmological framework consistent with current observations.
\subsection{Linear Matter Perturbations and Growth Evolution}
The present analysis focuses on the late-time cosmological evolution, where the anisotropic shear contribution becomes progressively subdominant. In this regime, the standard linear matter perturbation formalism of $f(T)$ gravity provides an effective approximation for studying the growth of large-scale structures. Accordingly, we employ the effective gravitational coupling $G_{\rm eff}=G/f_T$ to investigate the linear growth evolution of the proposed exponential $f(T)$ model. To further examine the cosmological viability of the model, we consider the evolution of linear matter perturbations. The matter density contrast $\delta_m$ satisfies
\begin{equation}
	\ddot{\delta}_m+2H\dot{\delta}_m
	-4\pi G_{\rm eff}\rho_m\delta_m=0,
\end{equation}

where

\begin{equation}
	G_{\rm eff}=\frac{G}{f_T},
	\qquad
	f_T=1+\mu\nu e^{\nu T}.
\end{equation}

The growth factor and growth rate are defined as

\begin{equation}
	D(z)=\frac{\delta_m(z)}{\delta_m(0)},
	\qquad
	f(z)=\frac{d\ln\delta_m}{d\ln a},
	\end{equation}

while the observable quantity is

\begin{equation}
	f\sigma_8(z)=f(z)\sigma_8(z).
\end{equation}
\begin{figure*}[t]
	\centering
	\includegraphics[width=0.9\textwidth]{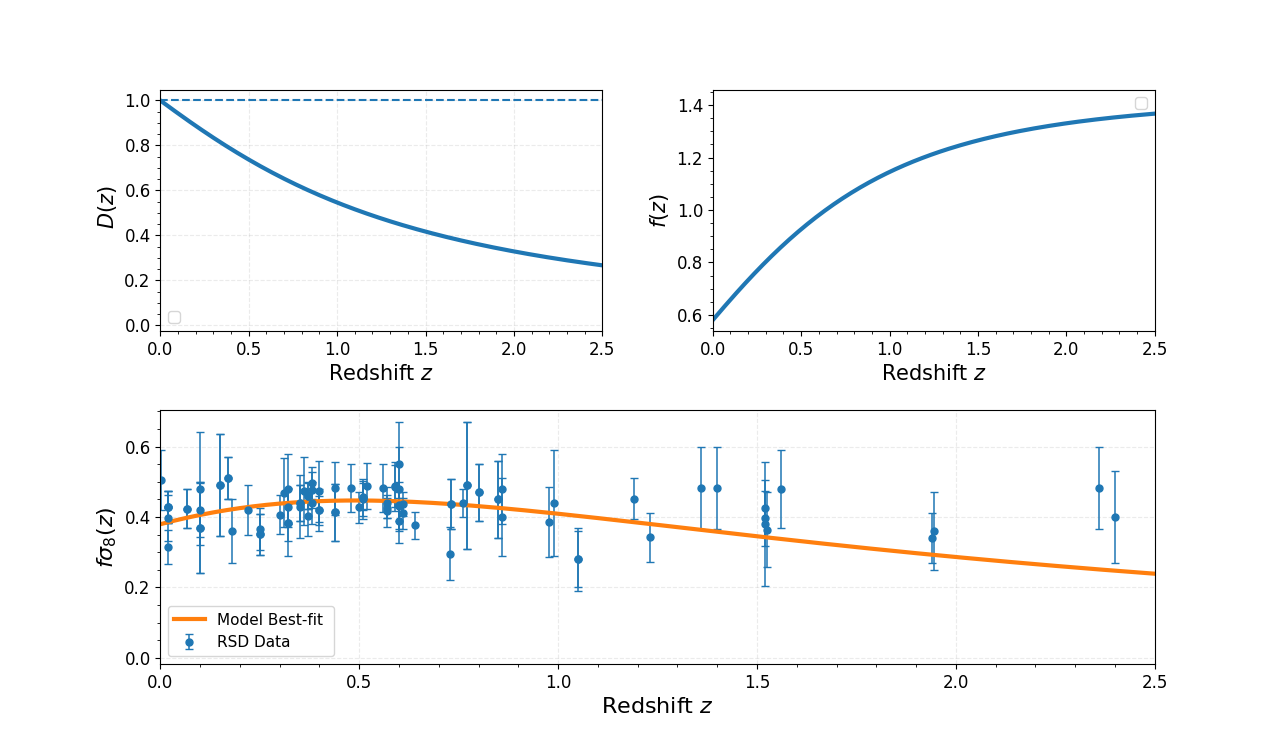}
	\caption{
		Behaviour of the linear growth observables in the exponential $f(T)$ cosmology. The panels show the evolution of the growth factor $D(z)$, growth rate $f(z)$, and the quantity $f\sigma_8(z)$ as functions of redshift. The observational growth data are included for comparison, illustrating the compatibility of the model with the observed structure formation history.
	}
	\label{fig:growth}
\end{figure*}
In Fig.~\ref{fig:growth}, we show the evolution of $D(z)$, $f(z)$, and $f\sigma_8(z)$. The growth factor evolves smoothly without instability, while the growth rate evolves smoothly with redshift and increases toward higher redshifts, approaching the expected matter-dominated growth behaviour. The predicted $f\sigma_8(z)$ evolution remains consistent with
the available observational data. The smooth behaviour of $D(z)$, $f(z)$, and $f\sigma_8(z)$ suggests that the model provides a consistent description of cosmic structure growth within the linear perturbation
regime.

\section{Conclusions}
The present accelerated expansion of the universe has been confirmed through several cosmological observations, leading to the development of various dark energy and modified gravity models to explain the late--time cosmic dynamics. Among these approaches, anisotropic cosmological models in the framework of $f(T)$ gravity have attracted considerable attention due to their ability to describe the accelerated expansion through torsional effects.

Motivated by these developments, the present work investigates a Bianchi type-I cosmological model in the framework of exponential $f(T)$ gravity with the functional form
$f(T)=T+\mu e^{\nu T}$.
The corresponding Hubble parameter is derived directly from the field equations, and the resulting model parameters were constrained using different combinations of observational datasets, including OHD, DESI DR2 BAO, Pantheon Plus, and SH0ES measurements. The obtained results have shown that the proposed model has successfully described the observed cosmic expansion and has provided a consistent description of the late--time accelerated universe within the modified teleparallel gravity framework.

The reconstructed Hubble parameter and distance modulus evolutions, shown in Figs.~\ref{fig:cc} and \ref{fig:sne}, have exhibited good agreement with the observational OHD, DESI DR2 BAO, Pantheon Plus, and SH0ES datasets over the considered redshift range. The obtained results have remained consistent with the standard $\Lambda$CDM cosmology and successfully reproduced the observed cosmic expansion behaviour. Furthermore, the statistical analysis presented in Tables~\ref{tab:chi} and \ref{tab:values} has supported the viability of the model, with the reduced $\chi^2$ values remaining close to unity. The corresponding AIC and BIC analyses have shown that the statistical comparison depends on the dataset combination. For the OHD+DESI DR2 BAO combination, the $\Lambda$CDM model yields lower AIC and BIC values, while the inclusion of shear reduces the minimum $\chi^2$ for the combined datasets. Moreover, the inclusion of shear contribution has introduced only minor deviations in the overall cosmological evolution. These results have indicated that the proposed anisotropic exponential $f(T)$ model with the exponential form given in Eq.~(\ref{g8}) has provided a reliable description of the observed expansion history of the universe.

The corner plot analysis shown in Fig.~\ref{fig:corner} has further illustrated the observational constraints on the model parameters and has shown that the obtained parameter spaces have remained well bounded for all dataset combinations. The corresponding confidence contours have demonstrated good consistency among the datasets and have supported the stability and observational viability of the proposed exponential $f(T)$ gravity model.
Table~\ref{tab:bestfit} presents the best-fit cosmological parameters obtained for the anisotropic exponential $f(T)$ model with and without shear contributions using different combinations of observational datasets. The estimated Hubble parameter values remain within the range
$69 \lesssim H_0 \lesssim 72~\mathrm{km\,s^{-1}Mpc^{-1}}$,
which is consistent with several recent observational constraints. In particular, TRGB measurements have suggested values around
$69$--$71~\mathrm{km\,s^{-1}Mpc^{-1}}$,
while combined BAO and supernova analyses have generally converged near
$70~\mathrm{km\,s^{-1}Mpc^{-1}}$.
Recent DESI observations have also reported values close to
$68$--$69~\mathrm{km\,s^{-1}Mpc^{-1}}$,
including
$H_0 = 68.86 \pm 0.68~\mathrm{km\,s^{-1}Mpc^{-1}}$
from WMAP+DESI analyses and
$H_0 = 68.72 \pm 0.51~\mathrm{km\,s^{-1}Mpc^{-1}}$
from WMAP+ACT+DESI observations. In contrast, Planck CMB observations have yielded  lower values around
$67$--$68~\mathrm{km\,s^{-1}Mpc^{-1}}$,
whereas SH0ES measurements have indicated relatively higher values near
$73$--$74~\mathrm{km\,s^{-1}Mpc^{-1}}$.
Therefore, the obtained Hubble parameter values remain well within the currently accepted observational bounds.

The matter density parameter is consistently constrained around
$\Omega_{m0}\simeq0.349$ for all the considered observational combinations, for both the shear and non-shear models. The inferred values are slightly higher than the matter density estimates commonly obtained from recent CMB and DESI-based analyses, which are typically around $\Omega_{m0}\sim0.31$. Nevertheless, the relatively tight constraints on $\Omega_{m0}$ across the different dataset combinations demonstrate the robustness of the parameter estimation within the present model. For the shear-inclusive cases, the present-day shear density parameter remains very small, with $\Omega_{\sigma0}$ lying in the range $\sim5.1\times10^{-7}$--$1.31\times10^{-6}$ across the considered combinations. These small values indicate that the anisotropic contribution is strongly subdominant at the present epoch, consistent with the observed near-isotropy of the late-time Universe. Thus, the model approaches an effectively isotropic regime at late times while retaining the possibility of a small anisotropic contribution during the cosmic evolution.

The evolution of the deceleration parameter, shown in Fig.~\ref{fig:dec}, indicates a clear transition from an earlier decelerated phase to the present accelerated expansion phase of the universe. The obtained present-day values remain negative for all observational dataset combinations, indicating the present accelerated expansion of the universe. Furthermore, the reconstructed trajectories show a qualitatively similar late-time evolution to the standard $\Lambda$CDM behaviour, although the transition redshifts vary depending on the observational dataset combination. The evolution of the effective equation of state parameter, shown in Fig.~\ref{fig:eos}, indicates that the obtained present-day values remain in the range approximately $-0.64\lesssim w^{\rm eff}_0\lesssim-0.45$ for all observational dataset combinations. At the present epoch, the reconstructed trajectories lie within the quintessence region, $-1<w^{\rm eff}_0<-1/3$, corresponding to the accelerated expansion phase. Towards higher redshifts, the trajectories cross the boundary $w^{\rm eff}=-1/3$, consistently indicating the transition from accelerated to decelerated expansion. Furthermore, the reconstructed $q(z)$ and $w^{\rm eff}(z)$ evolutions remain qualitatively consistent with the standard $\Lambda$CDM behaviour, supporting the capability of the proposed exponential $f(T)$ model to describe the observed late-time cosmic expansion.  

The reconstructed BAO observables shown in Figs.~\ref{fig:bao} and \ref{fig:three} remain consistent with the DESI DR2 BAO observations over the analysed redshift range. The obtained best-fit curves follow the observational data points reasonably well and show a similar evolution to the corresponding $\Lambda$CDM predictions. Furthermore, both the shear and non-shear cases reproduce the observed BAO distance measures with only small deviations, supporting the observational consistency of the proposed anisotropic exponential $f(T)$ gravity model with the large-scale expansion history of the universe. In addition, the reconstructed luminosity distance evolution, shown in Fig.~\ref{fig:gw}, remains compatible with the gravitational-wave observational data and exhibits a broadly consistent evolution with the standard $\Lambda$CDM behaviour over the considered redshift range.

The obtained results further support the viability of the exponential $f(T)$ functional form as a torsion-based framework for describing late-time cosmic acceleration. The observationally favoured parameter values correspond to very small values of the exponential parameter $\nu$, indicating that the exponential contribution introduces only mild torsional modifications during the late-time cosmological evolution. In this regime, the model exhibits stable cosmological behaviour, reproduces the observed accelerated expansion, and remains compatible with the considered observational datasets. By combining recent cosmological observations with dynamical, stability, and perturbation analyses, we show that the proposed model provides a consistent description of late-time cosmic evolution while supporting the isotropization of the universe at late times. These results highlight the relevance of torsional effects in describing the observed cosmic expansion within anisotropic exponential $f(T)$ gravity. The consistency between the reconstructed cosmological evolution, the stability requirements, and the perturbation behaviour provides additional support for the exponential form given in Eq.~(\ref{g8}) as a viable framework for describing the late-time universe. 

Overall, the present work demonstrates that the proposed anisotropic exponential $f(T)$ model, derived directly from the Bianchi type-I field equations, provides a physically viable and observationally consistent description of the late-time Universe. The combined observational constraints, dynamical system analysis, stability conditions, and linear matter perturbation analysis consistently support the robustness of the proposed framework. The statistical analysis further indicates that the shear-inclusive model provides an improved statistical fit for the combined observational datasets, as reflected by the comparatively lower $\chi^2$, AIC, and BIC values. Moreover, the best-fit results yield a very small but non-zero present-day shear contribution, indicating that the present Universe is close to isotropy while retaining a small residual anisotropic component. In three of the four combinations of observational datasets considered, the inferred shear contribution is particularly small, while the remaining combination also yields a small anisotropic contribution. These results are consistent with the expected tendency toward late-time isotropization within the proposed anisotropic exponential $f(T)$ framework. Thus, the statistical preference for the shear-inclusive model, together with the small present-day shear contribution, supports the picture of a nearly isotropic present Universe with a tendency toward late-time isotropization. Future investigations may extend the present analysis to alternative $f(T)$ functional forms and other anisotropic spacetime geometries. In addition, forthcoming high-precision observations from DESI \cite{DESI:2025zgx,DESI:2025zpo}, Euclid \cite{Euclid:2024yrr,Amendola:2016saw}, LSST \cite{LSST:2008ijt}, and next-generation gravitational-wave probes will allow tighter constraints on the model parameters. Furthermore, detailed investigations of cosmological perturbations, large-scale structure formation, and thermodynamical aspects within the torsion-based framework may provide deeper insight into the role of torsion in cosmic evolution.
	\section*{Acknowledgement}
	We gratefully acknowledge the use of data and resources provided by the Dark Energy Spectroscopic Instrument (DESI) collaboration. DESI is managed by the U.S. Department of Energy’s Office of Science and is supported by a collaboration of universities and laboratories worldwide. We also acknowledge the contributions of the Gravitational Wave Science Collaboration (GWSCO), whose efforts in advancing gravitational-wave detection and analysis have been invaluable to this work. Dr.~T.~Vinutha acknowledges the Inter-University Centre for Astronomy and Astrophysics (IUCAA), Pune, for providing research facilities and support under the Visiting Associateship Programme.The work of Kazuharu Bamba was supported in part by the JSPS KAKENHI Grants No. 24KF0100 and No. 25KF0176,  Competitive Research Funds for Fukushima University Faculty (25RK011), and a grant-in-aid of academic research of Yamaguchi Scholarship Foundation. Authors are very much thankful to the anonymous reviewer and editor for the valuable suggestions and constructive comments which have significantly improved the paper in terms of research quality and presentation.

\end{document}